\documentclass[twocolumn,fleqn,natbib]{svjour2}
\bibpunct{[}{]}{;}{n}{}{,} % to get "[numbered]" references from natbib
\smartqed  % flush right qed marks, e.g. at end of proof
\usepackage{comment}
\usepackage{epsfig}
\usepackage{amssymb}
\usepackage[cmex10]{amsmath}
\usepackage{amsfonts}
\usepackage{algorithm}
\usepackage{algorithmic}
\usepackage{mdwmath}
\usepackage{mdwtab}

\usepackage{graphicx}
\usepackage{float}
\usepackage{subfigure}

 \usepackage[titletoc]{appendix}

\newtheorem{lemma0}{\bf Lemma}

\newtheorem{definition0}{\bf Definition}

\newtheorem{theorem0}{\bf Theorem}

\journalname{Granular Matter}
\begin{document}

\title{Service-Constraint Based Truthful Incentive Mechanisms for Crowd Sensing%\thanks{Grants or other notes
}


\author{Jiajun Sun
}


\institute{J. Sun \at
              Beijing Key Lab of Intelligent Telecommunication Software and Multimedia, Beijing University of Posts and Telecommunications. \\
%              Fax: +123-45-678910\\
              \email{jiajunsun.bupt@gmail.com, mhd@bupt.edu.cn}           %  \\
           %\and
%           HD. Ma and D. Zhao \at
%              Beijing Key Lab of Intelligent Telecommunication Software and Multimedia, Beijing University of Posts and
%              Telecommunications. \\
%              \email{mhd@bupt.edu.cn,zhaodong86@gmail.com}           %  \\
}

\date{Received: 2014-04-05 }
% The correct date will be entered by the editor

\maketitle

\begin{abstract}
Crowd sensing is a new paradigm which leverages the pervasive
smartphones to efficiently collect and upload sensing data, enabling
numerous novel applications. To achieve good service quality for a
crowd sensing application, incentive mechanisms are necessary for
attracting more user participation. Most of existing mechanisms
apply only for the budget-constraint scenario where the platform
(the crowd sensing organizer) has a budget limit. On the contrary,
we focus on a different scenario where the platform has a service
limit. Based on the offline and online auction model, we consider a
general problem: users submit their private profiles to the
platform, and the platform aims at selecting a subset of users
before a specified deadline for minimizing the total payment while a
specific service can be completed. Specially, we design offline and
online service-constraint incentive mechanisms for the case where
the value function of selected users is monotone submodular. The
mechanisms are individual rationality, task feasibility,
computational efficiency, truthfulness, consumer sovereignty,
constant frugality, and also performs well in practice. Finally, we
use extensive simulations to demonstrate the theoretical properties
of our mechanisms. 
\end{abstract}

\section{Introduction}\label{sec:intro}
Crowd sensing is a new paradigm, which utilizes pervasive
smartphones to efficiently collect and upload data. Nowadays, the
proliferation of smartphones makes it possible to provide a new
opportunity for extending from the virtual space (online social
networks) to a larger real physical world (Internet of Things),
making users' contributions easier and omnipresent, such as Nericell
\cite{mohan2008nericell}, SignalGruru
\cite{koukoumidis2011signalguru}, and VTrack
\cite{thiagarajan2009vtrack} for providing omnipresent traffic
information, Ear-Phone \cite{rana2010ear} and NoiseTube
\cite{maisonneuve2009noisetube} for making noise maps. %To achieve
%good service quality, one of the main challenge the platform (the
%provider of crowd sensing applications) faces is in providing
%appropriate incentives for users so as to guarantee adequate user
%participation.

While participating in these applications, smartphone users consume
their own resources such as battery and computing power, and
disclose their locations with potential privacy threats. Thus,
incentive mechanisms are necessary to provide participants with
enough rewards for their participation costs. There are several
incentive mechanism studies for guaranteeing adequate user
participation in past literature. Generally, two scenarios for these
incentive mechanisms were considered: the offline scenarios and
online scenarios. For example, for the offline scenarios, the
authors of \cite{yang2012crowdsourcingfang} designed truthful
incentive mechanisms for the user-centric model and platform-centric
model respectively. For the online scenarios, the authors of
\cite{badanidiyuru2012learning,singer2013pricing,singlaminimi13}
designed incentive mechanisms based on the bidding model and the
posted price model for the additive utility function and submodular
utility function respectively. However, all these works only applied
for the scenario with the budget constraint where the platform with
a fixed budget aims at maximizing the platform's utility (e.g., the
total value of the tasks completed by selected users).

However, when the platform has a service limit instead of a budget
limit, which indicates that the platform need to minimize the total
payment for completing the fixed service, these truthful incentive
mechanisms become infeasible. To address this problem, the authors
of \cite{horton2010algorithmic,singer2013pricing} investigate the
frugality of incentive mechanisms for the offline scenario, in which
all of participating users report their profiles, including the
tasks they can complete and the bids, to the platform in advance,
and then the platform selects a subset of users after collecting the
sensing profiles of all users to minimize its total payments under
the condition that the specific tasks can be completed. But these
mechanisms only apply for the linear value function of sensing
tasks.
%For more
%details on crowd sensing applications, we refer interested readers
%to several surveys
%\cite{chatzimilioudis2012crowdsourcing,lane2010survey,ganti2011mobile}.

%However, in most of crowd sensing applications, users always arrive
%in a random order and user availability changes over time. The
%online opportunistic arrival manner of users applies to many real
%scenarios. For example, some platforms like Ear-Phone
%\cite{rana2010ear} and NoiseTube \cite{maisonneuve2009noisetube}
%allocates tasks to the smartphone users for monitoring some spatial
%phenomenon, such as air quality or traffic, only when the users
%arrive at the area of interest opportunistically. Therefore, an
%online incentive mechanism is necessary to make irrevocable
%decisions on whether to accept a user's task and bid sequentially,
%based solely on the current information of users arriving without
%knowing future information. On the other hand, in most of the above
%real crowd sensing scenarios, the value function of users is
%monotone submodular instead of additive, thereby the task-constraint
%based incentive mechanism are not infeasible for achieving good
%service quality.

In this paper, we concern a more general case, where the value
function of selected users' services is monotone submodular for
service constraints, instead of additive function supporting the
homogeneous and heterogeneous tasks. We investigate the offline and
online scenarios respectively for monotone submodular for service
constraints. For the offline scenario, the platform procures a
optimal solution to a given sensing services while minimizing the
total payment at the end of a specified deadline. For online
scenario, where users always arrive in a sequential order, and user
availability changes over time, so as to apply to most of the above
crowd sensing applications, the platform online determines whether
to select a user for a given sensing services while minimizing the
total payment. For the two scenarios, we consider users who are
game-theoretic and seek to make strategy (possible report a false
cost or arrival/departure time) to maximize their individual utility
in equilibrium. Thus, the problem of selecting feasible users while
minimizing the total payment can be modeled as the offline and
online auctions under the service and time constraints.

%Our objective is to design online mechanisms satisfying the
%following desirable properties: individual rationality,
%computational efficiency, truthfulness, consumer sovereignty and
%constant frugality. Informally, individual rationality ensures each
%participating user has a non-negative utility, computational
%efficiency ensures the mechanism can run in real time, truthfulness
%ensures the participating users report their true costs
%(cost-truthfulness) and arrival/departure times (time-truthfulness),
%consumer sovereignty ensures each participating user has a chance to
%win the auction, and constant frugality guarantees that the
%mechanism is close to the optimal solution in terms of total payment
%in the offline scenario where all users' information are known a
%priori.

%The main idea behind our online mechanisms is to adopt a
%multiple-stage sampling-accepting process. At every stage the
%mechanism allocates tasks to a user only if its bid is not less than
%a certain bid threshold that has been computed using previous
%users’ information, and the number of tasks allocated for the
%current stage has not been achieved. Meanwhile, the user obtains a
%bid-independent payment. The bid threshold is computed in a manner
%that guarantees desirable performance properties of the mechanism.
For the offline scenario, we adopt a ``myopic'' way to select the
optimal users to minimize the total payment. As long as the utility
function satisfies the submoduarity, a natural diminishing returns
condition, the mechanism satisfy the following critical properties:
1) Computational Efficiency: the auction can determine the winners
and payments in polynomial time; 2) Individual Rationality: each
user can expect a non-negative utility by participating in the
auction; 3) Service Constraint: It ensures the the platform's
service constraint is not violated. In this paper, service
constraint requires the mechanism to satisfy: $V(S)=R$; 4)
Truthfulness: no mobile user can benefit from cheating about its
true valuation on its cost of participation. For the online
scenario, we apply a multiple-stage sampling-accepting process to
solicit bids from users. At every stage the mechanism allocates
sensing tasks to an arriving smartphone user only if his marginal
utility is not less than a certain threshold density that has been
computed using previous users' bids and profiles as the sample set
until the service is completed. The threshold density is calculated
in a manner that guarantees the above desirable performance
properties of the offline mechanism. Besides, the online mechanism
also satisfies Constant Frugality: The mechanisms have constant
frugality ratio, i.e., if it announces the fixed services with the
value $R$ in expectation while guaranteeing that the total payment
is no more than the minimum cost required to achieve $\gamma R$
services in the offline scenario. The main contributions of this
paper are summarized as follows:

\begin{itemize}
\item  We design a service-constraint offline and online incentive mechanisms to ensure the minimal payment of the
platform for performing the required services respectively.
\item We rigorously prove that these incentive mechanisms
are satisfying the above desirable performances. We also evaluate
the performance and validate their theoretical properties via
extensive simulations.
\end{itemize}

The rest of the paper is organized as follows. In Section
\ref{related}, we briefly discuss the related work and motivation.
In Section \ref{SystemModel}, we present our system model and our
design goals. In Section \ref{offline} and Section \ref{sequential},
we design two service-constraint based incentive mechanisms for the
offline and online scenario respectively, followed by the
performance evaluation in \ref{Experiment}. Finally, Section
\ref{Conclude} concludes remarks.

\section{Background and Related Work}~\label{related}
There are growing interest in investigating the incentives for users
in online crowd sensing applications. For examples, the authors of
\cite{mason2010financial,shaw2011designing} study other,
non-monetary incentives that could improve the quality of users'
performance. The authors of \cite{chen2010new} apply no regret
learning to better understand users' behavior and improve the
results of sensing information aggregation from crowds. In contrast,
the authors of \cite{tran2012efficient} study the money incentives
to maximize tasks by using bandit algorithms. While it is a natural
approach, they leave room for frameworks that allow better
theoretical guarantees as used in this paper. The authors of
\cite{ho2012online} study an orthogonal problem and present an
algorithmic framework for matching users with requesters based on
their skills in crowd sensing applications.

Based on these frameworks, there are two classes of different model
studied extensively. One is to design the budget-constraint truthful
incentive mechanisms for stimulating adequate users to participate
in crowd sensing applications. For example, the authors of
\cite{badanidiyuru2012learning,singer2013pricing,singlaminimi13,yang2012crowdsourcingfang}
designed truthful incentive mechanisms for the offline and online
scenarios for maximizing the platform's utility. But these works
fail to handle the incentive problem of extensive user participation
under the service constraint. The other is to design
service-constraint incentive mechanisms for soliciting users' true
costs. For example, the authors of
\cite{horton2010algorithmic,singer2013pricing} investigates the
frugality of incentive mechanisms for the offline scenario with the
homogeneous and heterogeneous tasks. But they do not propose
feasible truthful incentive mechanisms for minimizing the total
payment. %the authors of \cite{singer2013pricing} propose a threshold
%mechanism for $\mathcal{O}(1)$-competitive online incentive
%mechanism design to maximize the utilities.
On the contrary, in this paper, we are interested in studying
minimizing payment online incentive mechanisms under given service
constraint for the offline and online scenarios, where the value
function of selected users' services is monotone submodular for
service constraints.
\section{System Model and Problem Formulation}~\label{SystemModel}
\subsection{System Model}
\begin{figure}
\setlength{\abovecaptionskip}{0pt}
\setlength{\belowcaptionskip}{10pt} \centering
%\begin{minipage}[t]{0.4\linewidth}
\centering
\includegraphics[width=0.30\textwidth]{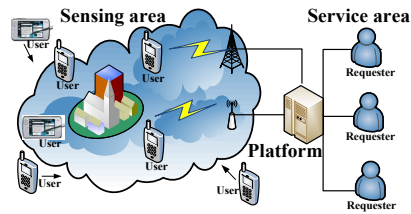}%[width=0.42\textwidth]{mrsu2.eps}%totalheight=3in,width=3.5in
\caption{Our crowd sensing system framework.} \label{crowd}
%\end{minipage}
\vspace{-10pt}
\end{figure}

We focus on crowd sensing applications with the goal to monitor some
spatial phenomenon, such as air quality or traffic. We consider the
following crowd sensing system model illustrated in
Fig.~\ref{crowd}. The system consists of a crowd sensing application
platform, which resides in the cloud and consists of multiple
sensing servers, and many mobile device users, which are connected
to the cloud by cellular networks (e.g., GSM/3G/4G) or WiFi
connections. The platform first publicizes a crowd sensing campaign
in an area of interest (AoI), aiming at finding some users to
complete a required utility value $R$ reflecting service quality
(given announced services). Then a set of users $\mathcal
{U}=\{1,2,\cdots,n\}$ interested in the campaign report their
profiles to the platform. Finally, the platform selects a feasible
subset of users to complete the given service before the deadline
$T$.

The platform is only interested in minimizing the total payment to
the selected users under the given service limit. We denote the
total services of the campaign as a finite set of locations,
$\Gamma=\{\tau_{1},\tau_{2},\cdots,\tau_{m}\}$, where each
$\tau_{i}\in \Gamma$ could, e.g., denote a zip code or more fine
grained street address, depending on the crowd sensing application.
Each user can sense a subset $\Gamma_{i}$ of $\Gamma$
($\Gamma_{i}\subseteq\Gamma$) like the number of locations depending
on her geolocation or mobility as well as the type of device used,
and have the cost $c_{i}$ corresponding to $\Gamma_{i}$. All these
information form the profile of user $i$, i.e.,
$\mathcal{P}_{i}=(c_{i},\Gamma_{i})$. Since smartphones are owned by
different users, it is reasonable to assume that users are selfish
but rational. Hence each user only wants to maximize its own
utility, and will not participate in the campaign unless there is
sufficient incentive.

In this paper, we study two scenarios: the offline scenario and
online scenario, where the value function of selected users'
services is monotone submodular for service constraints. In the
offline scenario, all of participating users report their profiles
to the platform synchronously, and then the platform allocates
services to a subset of users by considering the profiles of all
users at once. Different from the batched and synchronized manner in
the offline scenario, the interactive process in the online scenario
is sequential and asynchronous. Each user arrives in a sequential
order and submits its profile. Receiving the profile, the platform
must make an irrevocable decision about how much payment to pay to
each arrival user before the user departs until reaching the service
quality required. We assume that in each time step, a single user
appears and the platform makes a decision that is based on the
information it has about the user and the history of the previous
$i-1$ stages. Generally, there are three classes of user models:
\textit{the i.i.d. model}, \textit{the secretary model}, and
\textit{the adversarial model}. The first model means that at each
time step the costs and values of users are drawn from some unknown
distributions. The second model means that the users' costs are
chosen by an adversary, however their arrival order is a permutation
that is drawn uniformly at random from the set of all possible
permutations. In the third model, the users' costs and their arrival
order are chosen by an adversary. Note that in the third model,
although the adversary cannot observe the actions the mechanism
takes, since it has full knowledge, the adversary chooses the worst
arrival order and costs. Thereby, the mechanism cannot obtain the
optimal solutions. Thus, in this paper, we only account for the two
models with respect to the distribution of users, described in
increasing order of generality: the i.i.d.model and the secretary
model.

\subsection{Problem Formulation}
We model the above service-constraint based interactive process
between the platform and users as an auction with service and time
constraints. Receiving the crowd sensing campaign from the platform,
each user $i$ provides its profile
$\mathcal{P}_{i}=(c_{i},\Gamma_{i})$ to the platform so as to expect
a payment in return for its service. Since we assume that users are
game-theoretic and seek to make strategy to maximize their
individual utility in equilibrium. Note that in its profile, only
its service $\Gamma_{i}$ is true so that the platform can identify
whether the given services are fulfilled. That is, user $i$ can
misreport his cost, since his cost is private and only known to
himself. Thus, our strategy space can allow user $i$ to declare
$\hat{\mathcal{P}}_{i}=(b_{i},\Gamma_{i})$, where $b_{i}$ is a
reserve price or a bid made by user $i$ so as to sell its service.
Assume that the platform has given announced services denoted as a
utility value $U_{0}$ that it is willing to achieve. In order to
complete the required sensing services, more formally, an
offline/online mechanism $\mathcal{M}=(f,p)$, which consists of an
allocation function $f: \mathcal{R}_{+}^{n}\rightarrow 2^{[n]}$ and
a payment function $p:
\mathcal{R}_{+}^{n}\rightarrow\mathcal{R}_{+}^{n}$, is needed. That
is, for users'
$\hat{\mathcal{P}}=(\hat{\mathcal{P}}_{1},\hat{\mathcal{P}}_{2},\cdots,\hat{\mathcal{P}}_{n})$,
the allocation function computes an allocation of services for a
feasible subset of users $\mathcal{S}\subseteq \mathcal{U}$ and the
payment function returns a payment vector to feasible users. Thus,
the utility of user $i$ is $p_{i}-c_{i}$ if it is selected, $0$
otherwise. The platform expect to minimize the payments while
achieving the quality of announced services, i.e.,

 \vspace{-5pt}
\begin{equation*}
\min \sum_{i\in\mathcal{S}}p_{i} ~~~~Subject ~to ~V(S)\geq R
\end{equation*}
where $V(S)$ is the monotone submodular value function of services
from the selected users $\mathcal{S}$, illustrated in the following
definition.

\begin{definition0}[\textbf{Submodular Function}]\label{df:submodule}
Let $\mathbb{N}$ be a finite set, a function $V$ :
$2^{\Omega}\rightarrow \mathcal{R}$ is submodular if
$V(S\cup\{i\})-V(S)\geq V(T\cup\{i\})-V(T), \forall S\subseteq
T\subseteq \Omega$, where $\mathcal{R}$ is the set of reals.
\end{definition0}

\section{Offline Mechanism under the Service Constraint} \label{offline}
In this section, we present an offline mechanism under the service
constraint, satisfying the previous desirable properties.

For crowd sensing applications in the offline scenario, the authors
of
\cite{singer2010budget,tran2012efficient,yang2012crowdsourcingfang}
apply the proportional share allocation rule proposed in
\cite{singer2010budget} to address the extensive user participation
issue. However, the mechanism only applies for the offline scenario
with the budget constraint. To address this problem, we present a
service-constraint offline incentive mechanism that satisfies the
previous desirable properties. Illustrated in Algorithm
\ref{alg:offline}, our mechanism consists of two phases: the winner
selection phase and the payment determination phase.

\begin{algorithm}[htb] %算法的开始
\renewcommand{\algorithmicrequire}{\textbf{Input:}}
\renewcommand\algorithmicensure {\textbf{Output:} }
\caption{OMS// An Offline Mechanism for the Service constraint} %算法的标题
\label{alg:offline} %给算法一个标签，这样方便在文中对算法的引用
\begin{algorithmic}[1] %这个1 表示每一行都显示数字
\REQUIRE User set $\mathcal{U}$, the service
constraint $R$. \\% and the positive reserve $m$ \\%, the period deadline $T$ \\
\ENSURE The set of winners  $S$. \\
// Phase 1: Winner selection under services $R$ \\
\STATE $S\leftarrow \emptyset$; $i\leftarrow\arg\max_{j\in\mathcal
{U}}V_{j}(S)/b_{j}$;\\

\WHILE{$V(S)< R$}
                    \STATE $S\leftarrow S\cup i$;\\
                    %\STATE Put $\mathcal {A^{'}}$ into the TA;\\
                    \STATE $i\leftarrow\arg\max_{j\in\mathcal {U}\setminus S}(V_{j}(S)/b_{j})$;\\
            \ENDWHILE
\STATE $B\leftarrow\sum_{j\in S}b_{j}$; \label{sixline}\\
// Phase 2: Winner selection under budget $B$ \\
\STATE $S\leftarrow \emptyset$; $i\leftarrow\arg\max_{j\in\mathcal
{U}}V_{j}(S)/b_{j}$;\\

\WHILE{$V_{i}(S)/b_{i}\geq V(S\cup i)/B$}
                    \STATE $S\leftarrow S\cup i$;\\
                    %\STATE Put $\mathcal {A^{'}}$ into the TA;\\
                    \STATE $i\leftarrow\arg\max_{j\in\mathcal {U}\setminus S}(V_{j}(S)/b_{j})$;\\
            \ENDWHILE

// Phase 3: Payment determination \\
\FOR{each user $i\in \mathcal{U}$} \STATE $p_{i}\leftarrow 0$;\\
\ENDFOR
\FOR{each user $i\in S$} \STATE $\mathcal{U}^{'}\leftarrow \mathcal{U}\backslash\{i\}$; $\mathcal{T}\leftarrow \emptyset$;\\
\REPEAT%\WHILE {$b_{i_{j}}\leq
%U_{i(j)}(\mathcal{T}_{j-1})B/U(\mathcal{T})$}
   \STATE $i_{j}\leftarrow \arg \max_{j\in \mathcal{U}^{'}\backslash \mathcal{T}}(V_{j}(\mathcal{T})/b_{j})$;\\
   \STATE $p_{i}\leftarrow \max \{p_{i}, \min \{b_{i(j)}, \eta_{i(j)}\}\}$;\\
   \STATE $\mathcal{T}_{j-1}\leftarrow \mathcal{T}$;
   $\mathcal{T}\leftarrow \mathcal{T}\cup \{i_{j}\}$;\\
\UNTIL {$V(\mathcal{T})\geq R$}
%\ENDWHILE
    \ENDFOR
 \RETURN ($S$, $p$);\\

\end{algorithmic}
\end{algorithm}

%\begin{definition0}[\textbf{Submodular Function}]\label{df:submodule}
%Let $\mathbb{N}$ be a finite set, a function $U$ :
%$2^{\Omega}\rightarrow \mathbb{R}$ is submodular if
%$U(S\cup\{i\})-U(S)\geq U(T\cup\{i\})-U(T), \forall S\subseteq
%T\subseteq \Omega$, where $\mathbb{R}$ is the set of reals.
%\end{definition0}

From Definition \ref{df:submodule}, we can know the utility function
$V$ is submodular and derive the following sorting according to
increasing marginal contributions relative to their bids from users'
set to find the largest $k$ satisfying $V(S\cup k)<R$.

\begin{equation}\label{marginaleq}
V_{1}/ b_{1}\geq V_{2}/ b_{2}\geq \cdots \geq V_{|\mathcal{U}|}/
b_{|\mathcal{U}|},
\end{equation}
where $V_{k}$ denotes $V_{k\mid S_{k-1}}$
($=V(S_{k-1}\cup\{k\})-V(S_{k-1})$), $S_{k}=\{1,2, \cdots,k\}$, and
$S_{0}=\emptyset$. %Furthermore, we can perform the winners'
%selection and make the corresponding payments from the platform.
To calculate the payment of each user, we sort the users in
$\mathcal{U}\backslash\{i\}$ similarly as follows:

\begin{equation}\label{marginaleq}
V_{i_{1}}(\mathcal{T}_{0})/ b_{i_{1}}\geq
V_{i_{2}}(\mathcal{T}_{1})/ b_{i_{2}}\geq \cdots \geq
V_{i_{n-1}}(\mathcal{T}_{n-2})/ b_{i_{n-1}},
\end{equation}

The marginal value of user $i$ at the position $j$ is
$BV_{i(j)}(\mathcal{T}_{j-1})/V(\mathcal{T}_{j})$, where
$B=\sum_{j\in S}b_{j}$. Assume that $k^{'}$ to be the position of
the last user $i_{j}\in\mathcal{U}\backslash\{i\}$, such that
$V(\mathcal{T}_{j})<R$. To guarantee the truthfulness, each winner
should be given the payment of the critical value. This indicates
that user $i$ can not win the auction if it reports higher than this
critical value. More details are given in Algorithm
\ref{alg:offline}, where
$b_{i(j)}=V_{i(j)}(\mathcal{T}_{j-1})b_{i_{j}}/V_{i_{j}}(\mathcal{T}_{j-1})$
and
$\eta_{i(j)}=V_{i(j)}(\mathcal{T}_{j-1})B/V(\mathcal{T}_{j-1}\cup\{i\})$.

Since the OMS mechanism is very similar with MSensing in
\cite{singer2010budget,yang2012crowdsourcingfang}, only with three
differences. The one is that the services allocated to the winners
is a constraint instead of a factor in the objective function. The
second one is that OMS is a frugal mechanism instead of a budget
constraint mechanism, hence introducing line \ref{sixline} of
Algorithm \ref{alg:offline}. But these lines' introduction has no
impact on the following desirable properties. Thus, putting these
together, we have the following theorem.

\begin{theorem0}\label{theo:offine}
The OMS mechanism satisfies individual rationality, computational
efficiency, service feasibility, and truthfulness under the offline
scenario.
\end{theorem0}

\section{Online Mechanism under the Service Constraint} \label{sequential}

In this section, we present an online mechanism for the
service-constraint online scenario, satisfying all desirable
properties. To facilitate understanding, it is also assumed that
users arrive in a sequential order. But our mechanism can easily
apply generally or be extended to an random online scenario.

\subsection{Service-Constraint Online Mechanism Design}
%\subsection{Mechanism Design}
%When a user arrives, the platform
%must determine whether to buy his service, and if so, at what price
%before he departs.
%we are interested in
%mechanisms that implement an adaptive policy taking into account the
%observations made so far (revealed sensing profiles of users already
%selected) when choosing the next user, instead of committing to a
%fixed set of participants S in advance (non-adaptive policy).
An online mechanism needs to overcome several nontrivial challenges.
First, the users' costs are unknown and need to be elicited in a
truthful reporting manner. Second, an announced services should be
completed before the deadline. Finally, the mechanism needs to
tackle the online arrival of the users. To achieve good frugality,
previous online solutions and generalized secretary problems
\cite{hajiaghayi2004adaptive,bateni2010submodular,singer2013pricing,kleinberg2005multiple}
is via sampling: the first batch of the input is rejected and used
as a sample which enables making an informed decision on the rest of
the users. Since users are likely to be discouraged to sense data
knowing the pricing mechanism will automatically reject their bid.
In other words, those users arriving early have no incentive to
report their bids to the platform, which may delay the users'
completion or even lead to task starvation, i.e., the consumer
sovereignty issue in economics. Although the author of
\cite{kleinberg2005multiple} adopts a multi-stage sampling-accepting
process, it applies Dynkin's algorithm \cite{dynkin1963optimum} for
the classic secretary problem at the initial stage. Obviously, this
solution also cannot ensure the above task-starvation issue, since
Dynkin's algorithm adopts a two-stage sampling-accepting process.

To address the above challenges, we introduce a multi-stage
sampling-accepting process to design our online incentive mechanism.
At each stage, based on the above submodularity, the mechanism
maintains a density thres-\\hold which is used to decide whether to
accept the users' bids. The mechanism dynamically increases the
sample size and learns a budget that are enough to allocate users
for fulfilling the required services, then apply this budget to
compute a density threshold by applying budget feasible mechanisms,
and finally apply this density threshold for making further
decisions.

\begin{figure}
\setlength{\abovecaptionskip}{0pt}
\setlength{\belowcaptionskip}{10pt} \centering
%\begin{minipage}[t]{0.4\linewidth}
%\centering
\includegraphics[width=0.40\textwidth]{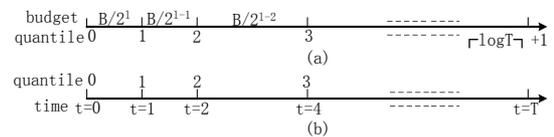}%[width=0.42\textwidth]{mrsu2.eps}%totalheight=3in,width=3.5in
\caption{Illustration of a multi-stage sample process with deadlines
$T$. (a)Budget constraints over quantiles; (b)Quantiles over
quantiles.} \label{multistage} %\vspace{-0.18in}
\vspace{-10pt}
\end{figure}

Specifically, our mechanism (see Algorithm \ref{sosstructure})
iterates over $q_{i}\in\{0,1,\cdots,\lceil\log T\rceil\}$ and at
every time step $q_{i}$, a required stage-service of $R^{'}=R/2^{i}$
is applied to allocate sensing services (illustrated in
Fig.~\ref{multistage}). This means that $R^{'}$ services should be
allocated before the end of this stage. Finally, the required
services $R$ should be allocated before the end of the deadline $T$.
At the beginning of the mechanism, we introduce a small value
$\varepsilon$ as initial density threshold. We assume that the
marginal value of user $i$ ($i\notin$) is $V_{i}(S)=V(S\cup\{i\})$,
where $S$ is selected users' set. In the sequel, as long as the
arrival user's marginal density $\frac{V_{i}(S)}{b_{i}}$ is not less
than the current threshold density value $\rho^{*}$ and the budget
has not been exhausted, the mechanism allocates service to it.
Meanwhile, we give user $i$ a payment $V_{i}(S)/\rho^{*}$, and add
this user to the set of selected users $S$.

\begin{algorithm}[htb] %算法的开始
\renewcommand{\algorithmicrequire}{\textbf{Input:}}
\renewcommand\algorithmicensure {\textbf{Output:} }
\caption{SOS// Service-constraint Online incentive mechanism
under a Sequential arrival model} %算法的标题
\label{sosstructure} %给算法一个标签，这样方便在文中对算法的引用
\begin{algorithmic}[1] %这个1 表示每一行都显示数字
\REQUIRE Service constraint $R$, sensing task deadlines $T$ \\
\STATE
$(t,T^{'},R^{'},S^{'},\rho^{*},S)\leftarrow(1,\frac{T}{2^{\lfloor\log_{2}T\rfloor}},\frac{R}{2^{\lfloor\log_{2}T\rfloor}},\emptyset,\varepsilon,\emptyset)$;\\

\FOR{$t\leq T$}
            \IF{there is a user $i$ arriving at time step $t$}\label{threeline1}
                \IF {$b_{i}\leq V_{i}(S)/\rho^{*}$ and $V(S)<R^{'}$ }\label{fourline1}
                    \STATE $p_{i}\leftarrow V _{i}(S)/\rho^{*}$, $S=S\cup\{i\}$;\label{fiveline1}\\

                \ELSE \STATE$p_{i}\leftarrow 0$;\label{sixline1}\\
                \ENDIF
                \STATE $S^{'}\leftarrow S^{'}\cup\{i\}$;\\
            \ENDIF\label{eightline1}
            \IF{$t=\lfloor T^{'}\rfloor$}
                \STATE Calculate $\rho^{*}\leftarrow$ getDensityThreshold($R^{'}$, $S^{'}$);\\
                \STATE set $R^{'}\leftarrow 2R^{'}$, $T^{'}\leftarrow 2T^{'}$;\\
            \ENDIF
            \STATE $t\leftarrow t+1$;\label{thirteenline1}\\
\ENDFOR
\end{algorithmic}
\end{algorithm}

In the computation of the density threshold for the mechanism, we
first find the maximal density for fulfilling $\delta R^{'}$
services from the sample set $S^{'}$. Then the process is repeated
by using a simple greedy manner until all of $\delta R^{'}$ services
are allocated. The greedy manner sorts users according to their
density, preferentially allocates services to users with higher
density. Here, we set $\delta$ to blow up the required stage
services so that the constant blowup services can be allocated at
the next stage. Furthermore, we compute the total payment for
fulfilling the constant blowup services. Futhermore, the algorithm
calls the following the budget feasible mechanism for submodular
function and then sets the density threshold to be $\rho/\nu$. $\nu$
is introduced to guarantee enough users selected and avoid the waste
of payment.

The above budget feasible mechanism for submodular function is an
offline mechanism proposed in \cite{singer2010budget}. It adopts a
proportional share allocation rule \cite{singer2010budget} to
compute the density threshold from the sample set $S^{'}$ and the
budget $B^{'}$. First of all, users are sorted according to their
increasing marginal densities. In this sorting the $(i+1)$-th user
is the user $j$ such that $V_{j}(S_{i})/b_{j}$ is maximized over
$S^{'}\setminus S_{i}$, where $S_{i}=\{1,2,\cdots,i\}$ and
$S_{0}=\emptyset$. Considering the submodularity of $V$, this
sorting implies that $\frac{V_{1}(S_{0})}{b_{1}}\geq
\frac{V_{2}(S_{1})}{b_{2}}\geq \cdots \geq
\frac{V_{|S^{'}|}(S_{|S^{'}|-1})}{b_{|S^{'}|}}.$

Then, the computation process adopts a greedy strategy. That is,
according to increasing marginal contributions relative to their
bids from the sample set to find the largest $k$ satisfying $b_{k ^*
}  \le \frac{R^{'}V_{k}(S_{k-1})}{V(S_{k})}$. Furthermore, we can
obtain the payment threshold estimated based on every sample set
$S^{'}$ with the privacy profile of users and the allocated
stage-budget $R^{'}$. Finally, we set the density threshold to be
$\frac{V(S_{k})}{\delta R^{'}}$. The detailed computation of the
threshold density is illustrated in Algorithm \ref{DensityThreshold}
and Fig.~\ref{multistage}.

%\begin{figure}
%\setlength{\abovecaptionskip}{0pt}
%\setlength{\belowcaptionskip}{10pt} \centering
%%\begin{minipage}[t]{0.4\linewidth}
%\centering
%\includegraphics[width=0.30\textwidth]{privacyconcern.eps}%[width=0.42\textwidth]{mrsu2.eps}%totalheight=3in,width=3.5in
%%\renewcommand{\figurename}{fig2}
%\caption{Illustration of our privacy-respecting online incentive
%mechanism which interacts with users.} \label{online}
%%\end{minipage}
%%\vspace{-0.24in}
%\end{figure}

\begin{algorithm}[t] %算法的开始
\renewcommand{\algorithmicrequire}{\textbf{Input:}}
\renewcommand\algorithmicensure {\textbf{Output:} }
\caption{getDensityThreshold} %算法的标题
\label{DensityThreshold} %给算法一个标签，这样方便在文中对算法的引用
\begin{algorithmic}[1] %这个1 表示每一行都显示数字
\REQUIRE Sample user set $S^{'}$, the stage-service $R^{'}$.\\%, bids $\textbf{b}_{\mathcal {W}^{'}}$.\\ %reduced budget factor $\alpha$. \\% and the positive reserve $m$ \\%, the period deadline $T$ \\
\ENSURE The threshold density $\rho$. \\
%\STATE // Phase 1: Winner selection \\
\STATE Initialize: $\mathcal {J}^{'}\leftarrow \emptyset$;
$i\leftarrow\arg\max_{j\in S^{'}}\frac{V_{j}(\mathcal{J})}{b_{j}}$; \label{oneline2}\\

 \WHILE{$V(\mathcal{J})< \delta R^{'}$}
                    \STATE $\mathcal {J}\leftarrow \mathcal {J}\cup \{i\}$; \\
                    \STATE Compute $i\leftarrow\arg\max_{j\in S^{'}\setminus\mathcal {J}}\frac{V_{j}(\mathcal{J})}{b_{j}}$; \\
            \ENDWHILE
\STATE $B^{'}\leftarrow \sum_{j\in\mathcal{J}}b_{j}$;\label{sixline2}\\
\STATE $\rho\leftarrow$ getFeasibleDensity$(B^{'}, S^{'})$;\label{sevenline2}\\
\RETURN $\rho/\nu$;\\
\end{algorithmic}
\end{algorithm}

\begin{algorithm}[htb] %算法的开始
\renewcommand{\algorithmicrequire}{\textbf{Input:}}
\renewcommand\algorithmicensure {\textbf{Output:} }
\caption{getFeasibleDensity \cite{singer2010budget}} %算法的标题
\label{FeasibleDensity} %给算法一个标签，这样方便在文中对算法的引用
\begin{algorithmic}[1] %这个1 表示每一行都显示数字
\REQUIRE Sample user set $S^{'}$, the budget $B^{'}$.\\%, bids $\textbf{b}_{\mathcal {W}^{'}}$.\\ %reduced budget factor $\alpha$. \\% and the positive reserve $m$ \\%, the period deadline $T$ \\
\ENSURE The threshold density $\rho$. \\
%\STATE // Phase 1: Winner selection \\
\STATE Initialize: $\mathcal {J}^{'}\leftarrow \emptyset$;
$i\leftarrow\arg\max_{j\in S^{'}}\frac{V_{j}(\mathcal{J})}{b_{j}}$;\\

 \WHILE{$b_{i  }  \le \frac{B^{'}{V _{i
 }(\mathcal{J}) }}{V(\mathcal{J}\cup \{i \})}$  and $V(\mathcal
{J})\leq B^{'}$ }
                    \STATE $\mathcal {J}\leftarrow \mathcal {J}\cup \{i\}$; \\
                    \STATE Compute $i\leftarrow\arg\max_{j\in S^{'}\setminus\mathcal {J}}\frac{V_{j}(\mathcal{J})}{b_{j}}$; \\
            \ENDWHILE
\STATE $\rho\leftarrow V(\mathcal{J})/B^{'}$;\\
\RETURN $\rho$;\\
\end{algorithmic}
\end{algorithm}

%\subsection{Mechanism Analysis}\label{basictools}
We now prove that our mechanism satisfies the desirable properties
as follows:

\begin{lemma0}\label{sampleofeffort1cost}
The SOS mechanism is incentive compatible or truthful.
\end{lemma0}

\begin{proof}
To see that bid-independent auctions are truthful, here consider a
user $i$ with cost of $c_{i}$ that arrives at some stage for which
the threshold density was set to $\rho^{*}$. If by the time the user
arrives there are no remaining required stage services, then the
user's cost declaration will not affect the allocation of the
mechanism and thus cannot improve his utility by submitting a false
cost. Otherwise, assume there are remaining required stage services
by the time the user arrives. In case $c_{i}\leq V
_{i}(S)/\rho^{*}$, reporting any cost below $V _{i}(S)/\rho^{*}$
wouldn't make a difference in the user's allocation and payment and
his utility for each assignment would be $V
_{i}(S)/\rho^{*}-c_{i}\geq 0$. Declaring a cost above
$V_{i}(S)/\rho^{*}$ would make the user lose the auction, and his
utility would be $0$. In case $c_{i}
>V_{i}(S)/\rho^{*}$, declaring any cost above
$ _{i}(S)/\rho^{*}$ would leave the user unallocated with utility
$0$. If the user declares a cost lower than $V
_{i}(\mathcal{S})/\rho^{*}$ he will be allocated. In such a case,
however, his utility will be negative. Thus the user's utility is
always maximized by reporting his true cost: $b_{i}=c_{i}$. Putting
these discussions together, the SOS mechanism satisfies
bid-independence. According to Proposition 2.1 in
\cite{bar2002incentive}, i.e., if and only if an online auction is
bid-independent, it is truthful. Thus, Lemma
\ref{sampleofeffort1cost} holds.
\end{proof}

\begin{lemma0}\label{sampleofeffort2budget}
The SOS mechanism is service feasible.
\end{lemma0}

\begin{proof}
 At each stage $t\in\{0,1,\cdots,\lfloor\log_{2} T\rfloor,\lfloor\log_{2}
 T\rfloor+1\}$, the mechanism uses a stage-service of
 $R^{'}=\frac{2^{t-1}R}{2^{\lfloor\log_{2}T\rfloor}}$.
 From the lines \ref{fourline1}-\ref{fiveline1} of Algorithm
\ref{sosstructure}, we can see that it is guaranteed that the
current total allocated services does not exceed the stage-service
$R^{'}$. Specially, the service constraint of the last stage is $R$.
Therefore, every stage is service feasible, and when the deadline
$T$ arrives, the total allocated services does not exceed $R$. It is
possible that the total required services can not be fulfilled. To
the end, we compute the minimal cost for fulfilling a constant
blowup of the required services by a frugal ratio $\delta$ (see
Algorithm \ref{DensityThreshold}). As such, $R/2$ required services
could be allocated at the last stage while the total payment is no
more than the budget $B$. Thereby, the mechanism can guarantee that
each stage uses minimal payments to achieving the required stage
services by blowing up to $\delta R^{'}$ until the total required
services are fulfilled.  Thus, Lemma \ref{sampleofeffort2budget}
holds.
\end{proof}

\begin{lemma0}\label{sampleofeffort3efficient}
The SOS mechanism is computational efficient.
\end{lemma0}

\begin{proof}
Since the mechanism runs online, we only need to focus on the
computation complexity at each time step $t=\{1,2,\cdots,T\}$.
Computing the marginal value of user $i$ takes $O(\Gamma_{i})$ time,
which is at most $O(m)$. Thus, the running time of computing the
allocation and payment of user $i$ (lines
\ref{threeline1}-\ref{eightline1} of Algorithm \ref{sosstructure})
is bounded by $O(m)$. Next, we analyze the complexity of computing
the density threshold, namely Algorithm \ref{DensityThreshold}.
Finding the user with maximum marginal density takes $O(m|S^{'}|)$
time. Since there are $m$ tasks and each selected user should
contribute at least one new task, the number of winners is at most
$\min\{m, |S^{'}|\}$. Thus, the running time of lines
\ref{oneline2}-\ref{sixline2} of Algorithm \ref{DensityThreshold} is
bounded by $O(m|S^{'}|\min\{m, |S^{'}|\})$. The running time of line
\ref{sevenline2} of Algorithm \ref{DensityThreshold} is the same as
of lines \ref{oneline2}-\ref{sixline2} of Algorithm
\ref{DensityThreshold}. Thus, the computation complexity at each
time step (lines \ref{threeline1}-\ref{thirteenline1}) is bounded by
$O(m|S^{'}|\min\{m, |S^{'}|\})$. At the last stage, the sample set
$S^{'}$ has the maximum number of samples, being $n/2$ with high
probability. Thus, the computation complexity at each time step is
bounded by $O(mn\min\{m, n\})$. Thus, Lemma
\ref{sampleofeffort3efficient} holds.
\end{proof}

\begin{lemma0}\label{sampleofeffort51}
The SOS mechanism is individually rational.
\end{lemma0}
\begin{proof}
From the lines \ref{fourline1}-\ref{sixline1} of Algorithm
\ref{sosstructure}, we can see that $p_{i}\geq b_{i}$ if $i\in S$,
otherwise $p_{i}=0$. Therefore, we have individual gain
$u_{i}\geq0$. Thus, Lemma \ref{sampleofeffort51} holds.
\end{proof}

\begin{lemma0}\label{sampleofeffort5}
The SOS mechanism satisfies the consumer sovereignty.
\end{lemma0}
\begin{proof}
Each stage is an accepting process as well as a sampling process
ready for the next stage. As a result, users are not automatically
rejected during the sampling process, and are allocated as long as
their marginal densities are not less than the current threshold
density, and the allocated stage services has not been exhausted.
Thus, Lemma \ref{sampleofeffort5} holds.
\end{proof}

If the stage services could be achieved at each stage, then $R$
required services would be allocated finally. Since our SOS
mechanism consists of multiple stages, and dynamically increases the
stage services, it only needs to prove that $R/2$ required services
could be allocated at the last stage while the total payment is no
more than the budget $B$. Thereby, the mechanism can guarantee that
each stage uses minimal payments to achieving the required stage
services by blowing up to $\delta R^{'}$ until the total required
services are fulfilled. The frugality ratio for achieving the
required services would be $\delta$, since at the last stage the
budget $B$ is the minimal cost for fulfilling the required stage
services $\delta R^{'}=\delta R/2$ according to Algorithm
\ref{DensityThreshold}. The mechanism for minimizing payments is
originated from the observations that the stage-service constraint
at each stage can be changed into the budget constraint at the
correspondent stage. If we show that at least $R/2$ required
services could be allocated at the last stage under the budget
constraint $B$, then it is equivalent to that $R/2$ required
services could be allocated while the total payment is no more than
$B$. This means that the frugality ratio for achieving the required
services is $\delta$.

\begin{lemma0}\label{competitive}
The SOS mechanism satisfies $O(1)$-compet-\\itive, i.e., constant
frugal ratio. Specifically, under i.i.d. model, we can achieve the
announced services from the platform when the frugal ratio
$\delta=8$. Under the secretary model, we can achieve the announced
services from the platform when the frugal ratio $\delta=24$.
\end{lemma0}

The detailed proof is given in Appendix \ref{competitiveproof}. From
the above lemmas, the following theorem holds.

\begin{theorem0}\label{sampleofeffort10POZ}
The SOS mechanism satisfies computational efficiency, individual
rationality, service feasibility, truthfulness, consumer
sovereignty, and constant frugality under a sequential arrival
model.
\end{theorem0}

\section{Performance Evaluation}\label{Experiment}
To evaluate the performance of our service-constraint mechanisms, we
implemented the OMS and SOS mechanisms, and compared them against
the random mechanism, i.e., uses a simple greedy algorithm like
Algorithm \ref{DensityThreshold}, which adopts a naive strategy for
rewarding users based on an uninformed fixed bid threshold. The
performance metrics include the frugal ratio, the running time, and
the platform's value.

\subsection{Simulation Setup}

%We can estimate the parameters of the demand distribution and
%thereby obtain the expected total demand values $D(A)$ according to
%sensing data collected from MTurk. As an example, we provide air
%quality for each start and destination online using mobile sensors
%by applying the obfuscation technology. We consider a granularity
%level of zip codes and locations $\mathcal{V}$ correspond to the zip
%codes. We obtained information related to latitude, longitude, city
%and county of these zips from publicly available data
%\footnote{http://www.populardata.com/downloads.heml}. In order to
%estimate the demand model, we use 3166 route planning requests
%obtained from users of a context-sensitive routing prototype used by
%volunteers at Microsoft. To create privacy profiles, we fixed the
%privacy degree requirements to a constant for all users. We also
%considered obfuscation within a fixed radius, centered around the
%user's location. For each of the obfuscated zip codes, multiple
%corresponding sensing profiles are generated, which collectively
%define the user's privacy profile.

We set the deadline (T) to 1800s, and vary the required services (R)
from 200 to 2000 with the increment of 200. Users arrive according
to a Poisson process in time with arrival rate $\lambda$. We vary
$\lambda$ from 0.2 to 1 with the increment of 0.2. The sensing range
of each sensor is set to 7 meters. The cost of each user is
uniformly distributed over [1, 10]. The initial density threshold
($\epsilon$) of Algorithm 1 and 4 is set to 1. Note that this
threshold could be an empirical value for real applications. All the
simulations were run on a PC with 1.7 GHz CPU and 8 GB memory. Each
measurement is averaged over 100 instances. All the simulations were
run on a PC with 1.7 GHz CPU and 8 GB memory. Each measurement is
averaged over 100 instances.

\subsection{Evaluation Results}
We first evaluate the frugal ratio's impact on the OMS and SOS
mechanisms. Then when the frugal ratio is fixed, we evaluate their
performances against the random mechanism.

\begin{figure}[t]
\setlength{\abovecaptionskip}{0pt}
\setlength{\belowcaptionskip}{10pt} \centering
%\begin{minipage}[t]{0.4\linewidth}
\centering
\includegraphics[width=1.9in]{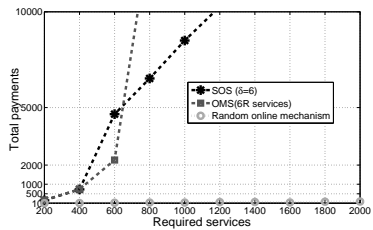}
%[width=0.42\textwidth]{mrsu2.eps}%totalheight=3in,width=3.5in
%\renewcommand{\figurename}{fig2}
\caption{Total payments versus required services.}
\label{paymentsfrugal}
%\end{minipage}
%\vspace{-0.24in}
\end{figure}

\noindent\textbf{Comparison on total payments:} The total payments
of all evaluated mechanisms increase with the value of required
services. From Fig.~\ref{paymentsfrugal}, we can observe that the
payments of the SOS mechanism ($\delta=6$) is lower than optimal
offline mechanism with $6R$ services. Note that, at most 4107
services can be completed by the OMS mechanism) in our simulations
due to the limit of the number of arrival users, and the $801$
services can be completed by the the SOS mechanism ($\delta=6$) and
the platform's payment is 4657.5 when the value of required services
is set as $800$, while the 3603 services can be completed by OMS,
and the platform's payment is 13940 under there are $6R=4801$
required services. The payment of SOS mechanism is much lower than
one of mechanism. This shows that the ``realistic'' frugality ratio
is less than $6$, which is consistent with our theoretical analysis
in Lemma \ref{competitive}. Thus, as the required services increase,
the mechanism SOS have lower payments than the OMS mechanism.
However, as the value of the required services increases, the
payments of the mechanism SOS are larger than the OMS mechanism. It
is because there is a limit of the number of available users in the
system. Additionally, Although Fig.~\ref{paymentsfrugal} shows that
random online mechanism has lower payments than our mechanisms, our
mechanisms ensure that required services are completed when there
are enough users to select. When the value of required services is
equal to $1200$, the services completed by the random online
mechanism are lower than half of required services, i.e. $591$.
%
%
%
%
%It is because that the optimal offline mechanism start to greedily
%select the users with maximal density, however our mechanisms start
%to select the users by setting a initial density, only when
%computing the density threshold, our mechanisms start to learn a
%optimal density. As the required services increase, our mechanisms
%have better performances than the optimal offline mechanism.
%Although random online mechanism has lower payments than our
%mechanisms, Fig.~\ref{completedservice} shows that completed
%services are much lower that required services. However, our
%mechanisms ensure that required services are completed when there
%are enough users to select. When the value of required services are
%more than the value $4000$, the difference between the completed
%services of our mechanisms and required services is produced due to
%the limitation of available users (here the number of available
%users is set as $1643$), as shown in Fig.~\ref{user}.

\begin{figure}[t]
\center \hspace{-0.08in} \subfigure[]{  \label{densitythreshold}
\includegraphics[width=1.6in]{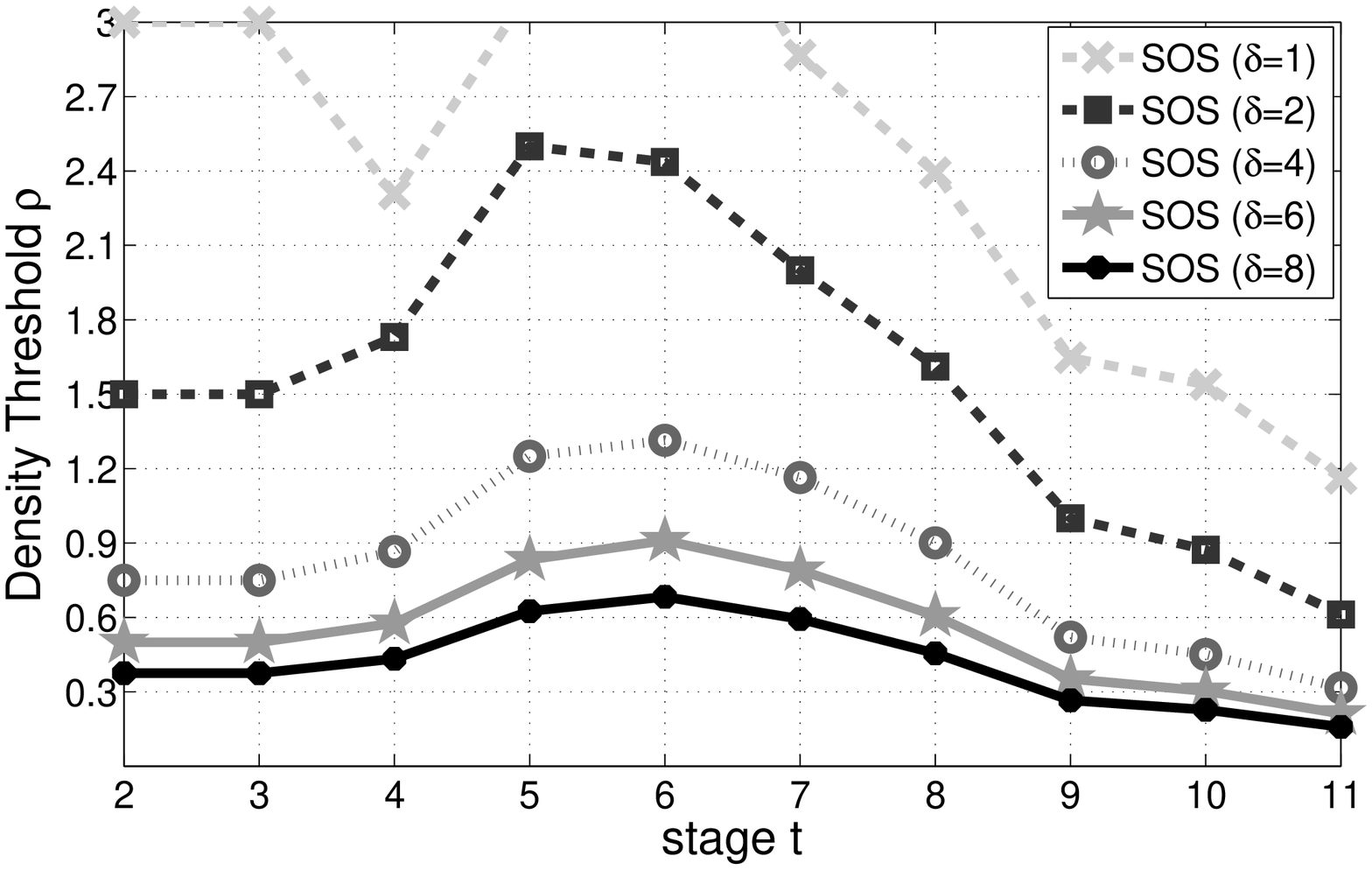}}
%\hspace{0.24in}
\subfigure[]{  \label{changefrugal}
\includegraphics[width=1.6in]{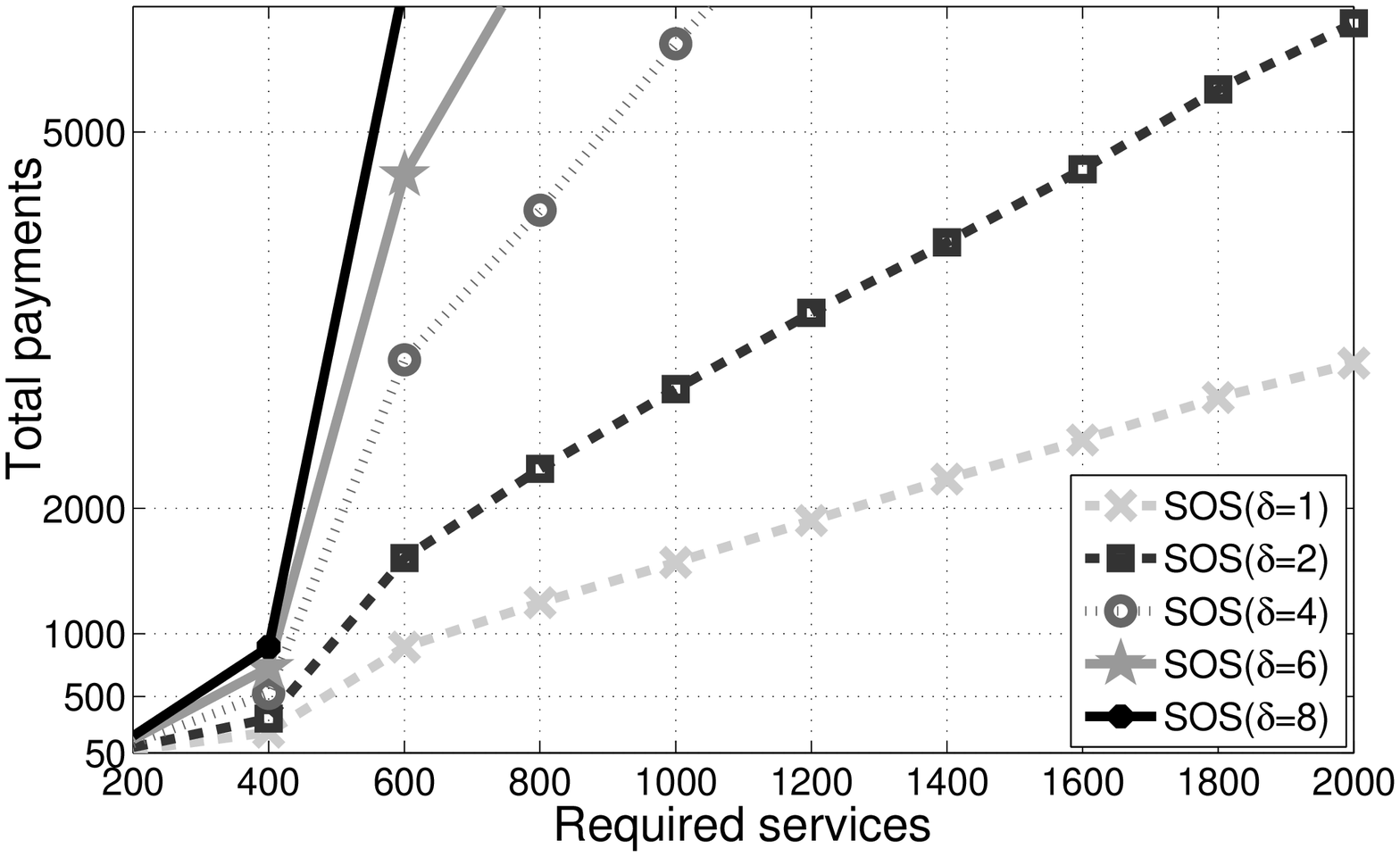}}
%\subfigure[]{ \label{completedservice}
%\includegraphics[width=1.6in]{completedservice.eps}}
%%\hspace{0.24in} \subfigure[]{ \label{valuebudget}
%%\includegraphics[width=2.4in]{valuebudget.eps}}
%\subfigure[]{ \label{user}
%\includegraphics[width=1.6in]{selecteduser.eps}}
\caption{ (a)Density threshold $\rho$ versus stage $t$ under
different frugal ratio $\delta$ when the value of required services
is set as $1000$; (b)The total payments versus required services
under different frugal ratio $\delta$.} \label{reportsandverify}
%\vspace{-20pt} ;(c)The number of selected users versus required services.
\vspace{-10pt}
\end{figure}

\noindent\textbf{Frugal ratio's impact:} Fig.~\ref{densitythreshold}
shows that the density threshold of each stage decreases as the
frugal ratio $\delta$ increases, thereby achieving much lower
payments. The density threshold of each stage tends to a constant
when the frugal ratio $\delta$ is larger than $8$. Thus, the SOS
mechanism learns a optimal density that achieves the minimal
payments meanwhile fulfilling required services. The SOS mechanism
can attain lower payment as the value of $\delta$ increases. When
the payments of the SOS mechanism are lower than the payments, we
call the value as the frugal ratio. Fig.~\ref{changefrugal} shows
that the total payments of the platform converges towards a constant
value with the increase of the frugal ratio $\delta$.
%the limitation of available users (here the number of available
%users is set as $1643$), as shown in Fig.~\ref{user}.
% compare that
%the fulfilled total services and the total payments of the platform
%changes as the size of frugal ratio $\delta$.

\begin{figure}[h]
\center \hspace{-0.08in} %\subfigure[]{  \label{runtime}
%\includegraphics[width=1.4in]{runtime.eps}}
%\hspace{0.24in}

\subfigure[]{ \label{completedservice}
\includegraphics[width=1.6in]{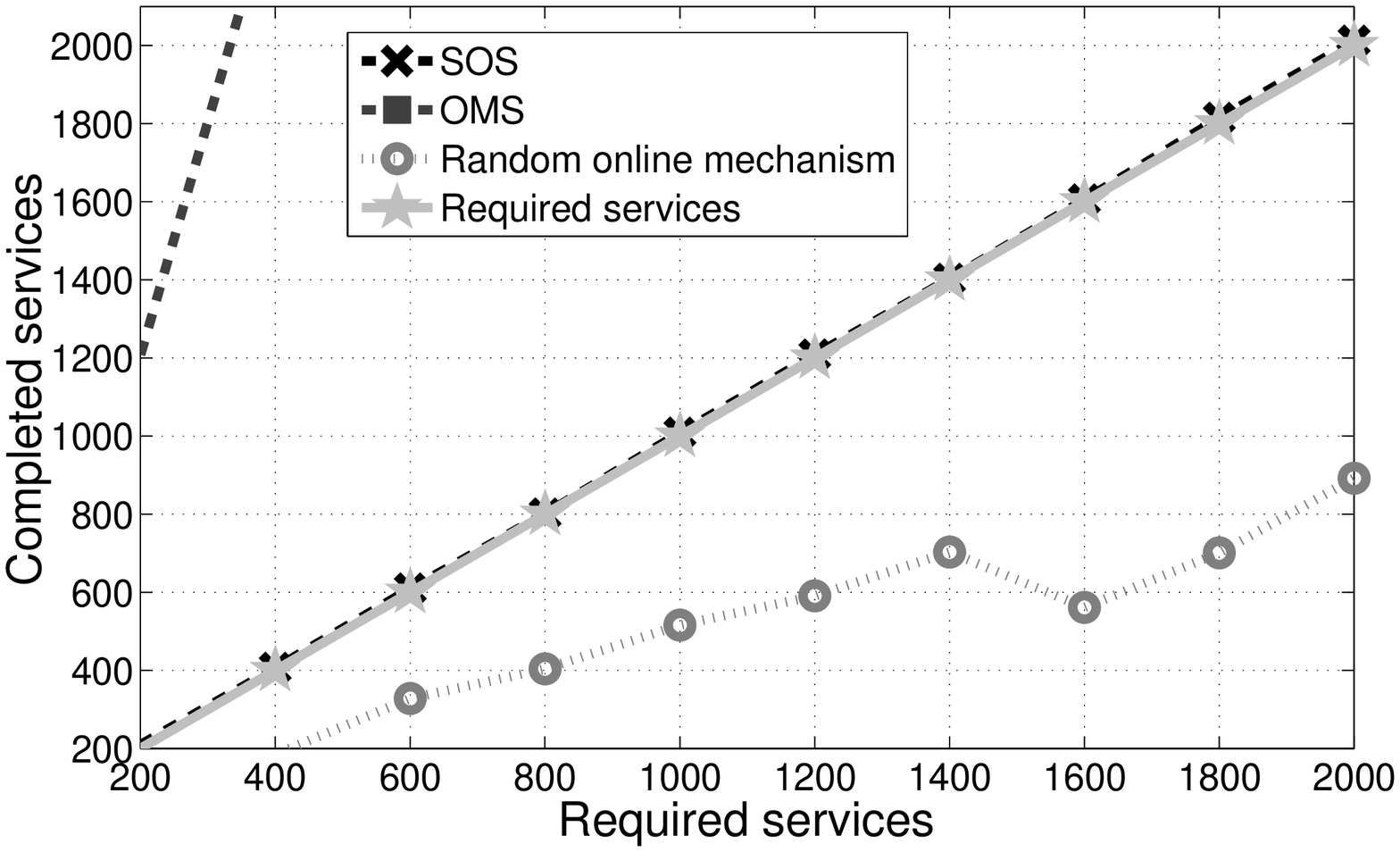}}
\hspace{0.14pt} %\subfigure[]{ \label{valuebudget}
\subfigure[]{ \label{user}
\includegraphics[width=1.6in]{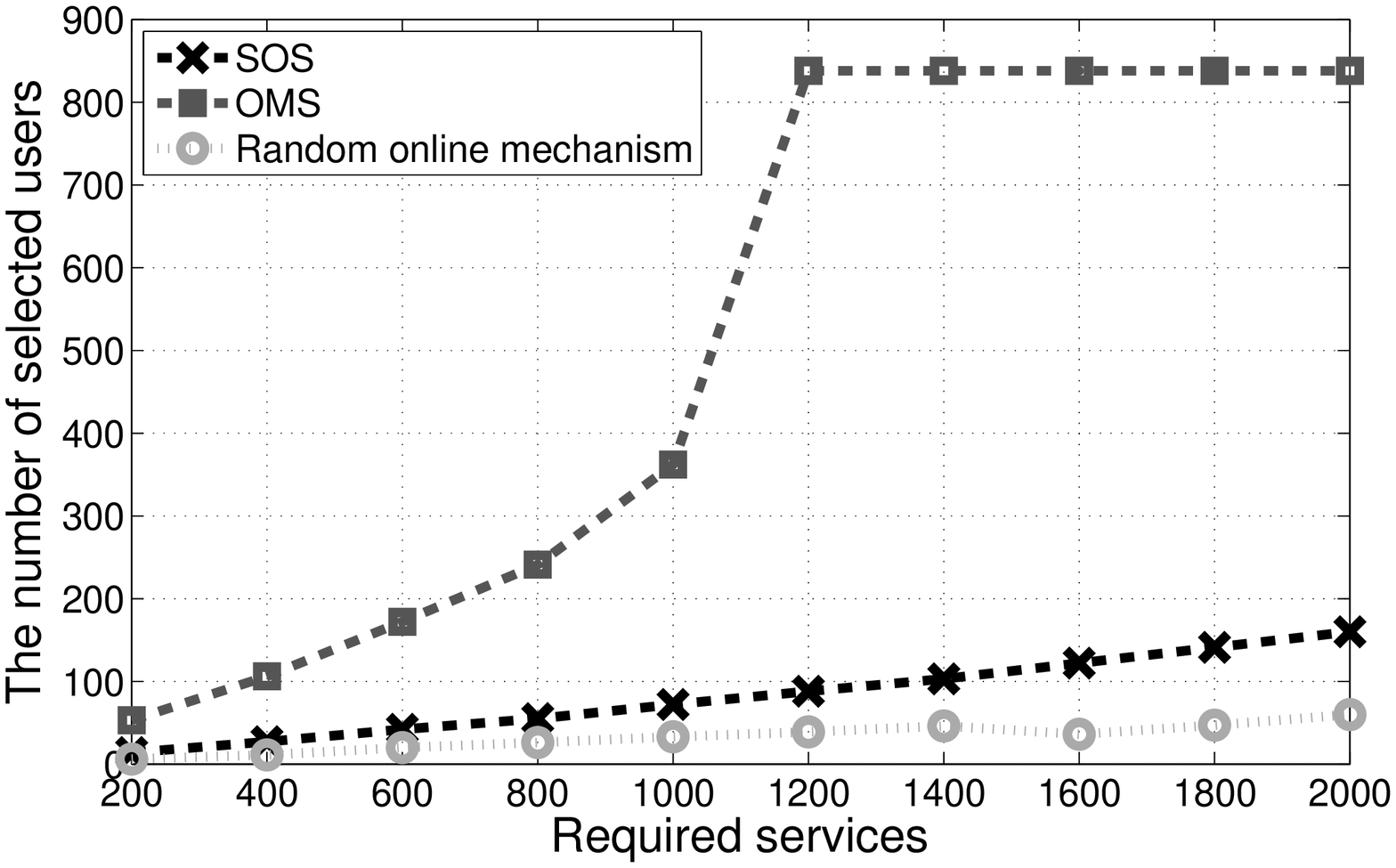}}
\caption{ (a)Completed services versus required services;(b) The
number of selected users versus required services when $\delta=6$.}
\label{reportsandverify} \vspace{-10pt}

\end{figure}

\noindent\textbf{Required service value's impact:}
Fig.~\ref{completedservice} shows that completed services of random
online mechanism are much lower that required services. However, the
SOS mechanism ensures that required services are completed when
there are enough users to select. When the value of required
services are more than the value $1000$, SOS mechanism completes all
required services while OMS mechanism does not complete $6R$
required services. It is because that there is the limitation of
available users in simulation (here the number of available users is
set as $838$). However, the limit of the number of available users
has no impact on the SOS mechanism, because in SOS mechanism only
need to select $163$ users to complete the required services $2000$,
as shown in Fig.~\ref{user}.

\vspace{-10pt}
\section{Conclusions}~\label{Conclude}
In this paper, we have designed two incentive mechanisms to motivate
smartphone users to participate in crowd sensing application with
the service constraint, which is a new sensing paradigm allowing us
to efficiently collect data for achieving required service quality.
We first propose a OMS mechanism for the offline scenario.
Furthermore, we design a SOS mechanism for a sequential arrival
model, where users arrive one by one online. We also prove that the
two mechanisms satisfy the above desirable properties.

%\begin{acknowledgements}
%This work is supported by the National Natural Science Foundation of
%China under Grant No.$61$\\$332005$, No.61272517, No.61133015, the
%Funds for Creative Research Groups of China under Grant No.61121001,
%the Cosponsored Project of Beijing Committee of Education, and the
%Key Technologies R\&D Program of China under Grant No.2011BAC12B03.
%\end{acknowledgements}

% BibTeX users please use
\bibliographystyle{spbasic}
\bibliography{IEEEtran}   % name your BibTeX data base

%\begin{thebibliography}{3}
%%
%% and use \bibitem to create references. Consult the Instructions
%% for authors for reference list style.
%\bibitem[Author I(1999)]{Ref1}
%Author I (year) Article title. Journal Title-Abbreviated Vol: pp--pp
%\bibitem[Author and Smith(2001)]{Ref2}
%Author I, Smith J (year) Book title. Publisher, Place, pp numbers
%% Format for proceedings
%\bibitem[Author and Smith(2003)]{Ref3}
%Author I, Smith J (year) Paper title. In: Editor, A. (ed.) Proceedings
%Title, Location, Date, pages. Publisher, Place
%\end{thebibliography}

\appendix

\renewcommand{\appendixname}{Appendix~\Alph{section}}

      \section{}
\noindent\textbf{{Proof of Lemma \ref{competitive}:}}\label{competitiveproof}\\
%According to Lemma 3.1 in \cite{singer2013pricing}, for a
Let $S^{*}$ be the set of users selected by the offline Algorithm
\ref{FeasibleDensity} before the time $T$ and the budget $2B$, the
value of $S^{*}$ is $V(S^{*})$. The value density threshold of
$S^{*}$ is $\rho=V(S^{*})/B$. $S^{'}$ is the sample set obtained at
the time $T/2$. $S_{1}^{*}=S^{*}\cap S^{'}$ and $S_{2}^{*}=S^{*}\cap
\{\mathcal{U}\setminus S^{'}\}$. $S_{1}^{'}$ is the set of users
selected from the sample set $S^{'}$ by Algorithm
\ref{FeasibleDensity} before the time $T$ and the budget $B$, and
$S_{2}^{'}$ is the set of users selected by Algorithm
\ref{sosstructure} at the last stage. Let
$\rho_{1}^{'}=V(S_{1}^{'})/B$ be the density computed using
Algorithm \ref{DensityThreshold} over $S^{'}$ and
$\rho^{*}=\rho_{1}^{'}/\nu$ is the density threshold of the last
stage. Assume that the value of each user is at most
$\max_{i}V_{i}\leq V(S^{*})/\omega$.

%In order to make Lemma \ref{competitive} hold, we first provide the
%following lemma.
%
%\begin{lemma0}\label{sampleofeffort9}
%Given a sample set $S^{'}$, the total value of selected users
%computed by Algorithm \ref{FeasibleDensity} with the budget
%$B^{'}/2$ is at least a half of that computed with the budget
%$B^{'}$.
%\end{lemma0}
%\begin{proof}
%Assume that the set of selected users computed with the budget
%$B^{'}/2$ is $S_{l}=\{1,2,\cdots,l\}$, and the set of selected users
%computed with the budget $B^{'}$ is $S_{k}=\{1,2,\cdots,k\}$. Then,
%users can be sorted according to their increasing margi-\\nal
%densities as follows: $V_{1}(S_{0})/ b_{1}\geq V_{2}(S_{0})/
%b_{2}\geq \cdots \geq V_{l}(S_{l-1})/ b_{l}\geq
%2V(S_{l})/\mathcal{B}^{'}\geq V_{l+1}(S_{l})/ b_{l+1}\geq \cdots
%\geq V_{k}(S_{k-1})/ b_{k}$\\$\geq V(S_{k})/B^{'}\geq
%V_{|S^{'}|}(S_{|S^{'}|-1})/ b_{|S^{'}|}$. Thus, it can be easily
%derived that: $V(S_{l})\geq V(S_{k})/2$. Thus, Lemma
%\ref{sampleofeffort9} holds.
%\end{proof}

\begin{proof}
In the proof, we consider that the mechanism is constant frugal from
the two class model: I.I.D. and the Secretary Model

Under I.I.D. Model, since the costs and values of all users in
$\mathcal{U}$ are i.i.d., they can be selected in the set $S^{*}$
with the same probability. Thus, we have
$\mathbb{E}[|S_{1}^{'}|]=\mathbb{E}[|S_{1}^{'}|]=|S^{*}|/2$.
Considering the submodularity of function $V(S)$, it can be derived
that: $\mathbb{E}[V(S_{1}^{*})]\geq\mathbb{E}[V(S_{2}^{*})]\geq
V(S^{*})/2=R/2$. Since $V(S_{1}^{'})$ is computed with the
stage-budget $B/2$, it can be derived that:
$\mathbb{E}[V(S_{1}^{'})]\geq\mathbb{E}[V(S_{1}^{*})]\geq
V(S^{*})/2=R/2$ and $\mathbb{E}[\rho_{1}^{'}]\geq \rho$. where the
first inequality follows from the fact that $V(S_{1}^{'})$ is the
optimal solution computed by Algorithm \ref{FeasibleDensity}.
Therefore, we only need to prove that the ratio of
$\mathbb{E}[V(S_{2}^{'})]$ to $\mathbb{E}[V(S_{1}^{'})]$ is at least
a constant, then the SOS mechanism have a constant frugal ratio.
Only two cases can exist according to the total payment to the
selected users at the last stage.

According to Lemma 7 in \cite{DBLPZhaoLM13}, we have
$1/2-(\frac{\nu}{1-2\alpha}-1)/\omega-1/\nu=2\alpha/\nu$. Thus, when
$\omega$ is sufficiently large (at least 12), we can obtain a
constant ratio of $\mathbb{E}[V(S_{2}^{'})]$ to
$\mathbb{E}[V(S_{1}^{'})]$. More importantly, the optimal ratio
increases to $1/4$ (i.e., $2\alpha/\nu\rightarrow1/4$) as $\omega$
increases.

From Lemma 9 in \cite{DBLPZhaoLM13}, we have
$\mathbb{E}[V(S_{1}^{'})]\geq \frac{\delta R}{4}$. Furthermore,
$\mathbb{E}[V(S_{2}^{'})]\geq\frac{2\alpha}{\nu}\mathbb{E}[V(S_{1}^{'})]\geq\frac{2\alpha}{\nu}\cdot\frac{\delta
R}{4}$. According to the previous discussions, to achieve the
required services, the inequality $\mathbb{E}[V(S_{2}^{'})]\geq R/2$
holds by setting $\frac{\alpha\delta R}{2\nu}\geq \frac{R}{2}$. As
such, we have $\delta\geq2\cdot\nu/2\alpha\geq2\times4=8$. Thus, we
can set the frugal ratio $\delta=8$ to achieve the required
services.

Under the Secretary Model, let $S^{*}$ be the set of users selected
by the offline Algorithm \ref{FeasibleDensity} before the time $T$
and the budget $B$ other than the budget $2B$ in the i.i.d. model.
According to Lemma 15 in \cite{bateni2010submodular}, for
sufficiently large $\omega$, the random variable $|V(S_{1}^{*}) -
V(S_{2}^{*})|$ is bounded by $V(S^{*})/2$ with a constant
probability. Because of the submodularity of $V$, we have
$V(S_{1}^{*}) +V(S_{2}^{*})\geq V(S^{*})$. Thus, we easily obtain
the result: For sufficiently large $\omega$, both $V(S_{1}^{*})$ and
$V(S_{2}^{*})$ are at least $V(S^{*})/4$ with a constant
probability. Putting the result and Lemma 9 in \cite{DBLPZhaoLM13}
together, we have $\geq V(S_{1})/2\geq V(S^{*})/8$ Only two cases
can exist according to the total payment to the selected users at
the last stage.

According to Lemma 10 in \cite{DBLPZhaoLM13}, we have
$1/4-(\frac{8\nu}{1-2\alpha}-1)/\omega-2/\nu=2\alpha/\nu$. Thus,
when $\omega$ is sufficiently large (at least 12), we can obtain a
constant ratio of $V(S_{2}^{'})$ to $V(S_{1}^{'})$. More
importantly, the optimal ratio increases to $1/12$ (i.e.,
$2\alpha/\nu\rightarrow1/12$) as $\omega$ increases.

In terms of Lemma 9 in \cite{DBLPZhaoLM13}, we have
$V(S_{1}^{'})\geq \frac{\delta R}{4}$. Furthermore,
$V(S_{2}^{'})\geq\frac{2\alpha}{\nu}V(S_{1}^{'})\geq\frac{2\alpha}{\nu}\cdot\frac{\delta
R}{4}$. According to the previous discussions, to achieve the
required services, the inequality $V(S_{2}^{'})\geq R/2$ holds by
setting $\frac{\alpha\delta R}{2\nu}\geq \frac{R}{2}$. As such, we
have $\delta\geq2\cdot\nu/2\alpha\geq2\times12=24$. Thus, we can set
the frugal ratio $\delta=24$ to achieve the required services.

Thus, the Lemma \ref{competitive} holds.
\end{proof}

\end{document}